\documentclass[aip,amsmath,amssymb,reprint,superscriptaddress]{revtex4-1}

\usepackage{graphicx}\usepackage{dcolumn}\usepackage{bm}

\usepackage[utf8]{inputenc}
\usepackage[T1]{fontenc}
\usepackage{mathptmx}
\usepackage{etoolbox}
\usepackage{array}
\usepackage{makecell}

\usepackage{algorithm2e}
\RestyleAlgo{ruled}
\SetKwComment{Comment}{// }{} 
\SetCommentSty{commentfont} \DontPrintSemicolon
\SetKwProg{Fn}{Function}{:}{}

\usepackage{color}
\usepackage{listings}
\usepackage{xcolor}

\definecolor{codegreen}{rgb}{0,0.6,0}
\definecolor{codegray}{rgb}{0.5,0.5,0.5}
\definecolor{codepurple}{rgb}{0.58,0,0.82}
\definecolor{backcolour}{rgb}{0.95,0.95,0.92}

\lstdefinestyle{mystyle}{
    backgroundcolor=\color{backcolour},   
    commentstyle=\color{codegreen},
    keywordstyle=\color{magenta},
    numberstyle=\tiny\color{codegray},
    stringstyle=\color{codepurple},
    basicstyle=\ttfamily\footnotesize,
    breakatwhitespace=false,         
    breaklines=true,                 
    captionpos=b,                    
    keepspaces=true,                 
    numbers=left,                    
    numbersep=5pt,                  
    showspaces=false,                
    showstringspaces=false,
    showtabs=false,                  
    tabsize=2
}
\newcommand{\st}[1]{_{\mathrm{#1}}}
\newcommand{\sr}[1]{^{\mathrm{#1}}}

\DeclareUnicodeCharacter{2212}{\textendash}
\DeclareUnicodeCharacter{226A}{$\ll$}

\usepackage{siunitx}
\makeatletter
\def\@email#1#2{\endgroup
 \patchcmd{\titleblock@produce}
  {\frontmatter@RRAPformat}
  {\frontmatter@RRAPformat{\produce@RRAP{*#1\href{mailto:#2}{#2}}}\frontmatter@RRAPformat}
  {}{}
}\makeatother
\begin{document}
\preprint{AIP/123-QED}

\title {\textit{mrfmsim}: a modular, extendable, and readable simulation package for magnetic resonance force microscopy experiments}
\author{Peter Sun}
\email{hs859@cornell.edu}
\affiliation{
  Department of Chemistry and Chemical Biology, Cornell University, 
  Ithaca, New York, 14853, USA
}
\author{Corinne E. Isaac}
\affiliation{
  Department of Chemistry and Chemical Biology, Cornell University, 
  Ithaca, New York, 14853, USA
}
\author{Michael C. Boucher}
\affiliation{
  Department of Chemistry and Chemical Biology, Cornell University, 
  Ithaca, New York, 14853, USA
}
\author{Eric W. Moore}
\affiliation{
  Department of Chemistry and Chemical Biology, Cornell University, 
  Ithaca, New York, 14853, USA
}
\author{Zhen Wang}
\affiliation{
  Department of Chemistry and Chemical Biology, Cornell University, 
  Ithaca, New York, 14853, USA
}
\author{John A. Marohn}
\affiliation{
  Department of Chemistry and Chemical Biology, Cornell University, 
  Ithaca, New York, 14853, USA
}

\begin{abstract}

We present \textit{mrfmsim}, an open-source Python package that facilitates the design, simulation, and analysis of magnetic resonance force microscopy (MRFM) experiments. MRFM is a scanning-probe technique that detects magnetic resonance from nanoscale ensembles of nuclear or electron spins with a force sensor. Because MRFM experiments are complex and operate at sensitivity limits, numerical simulation is essential for designing experiments and estimating per-spin sensitivity and imaging resolution from measured signals. In this paper, we highlight the challenges of developing MRFM simulations and show that software designed to simulate specific experiments only in a rapidly evolving experimental field can yield erroneous results. The \textit{mrfmsim} package addresses these challenges by supporting post-definition customization without rewriting the internal model and by employing a plugin system for extending functionality. We show that the package's modular, extendable, and readable architecture improves reproducibility and accelerates development.
 \end{abstract}

\maketitle

\section{Introduction}

Magnetic resonance force microscopy (MRFM) is a scanning probe technique that has detected and imaged magnetic resonance from an individual subsurface electron spin,\citep{Rugar2004jul} imaged nuclear density in a single copy of a virus,\citep{Degen2009feb} detected spin polarization,\citep{Klein2000aug, Thurber2002mar} and observed ferromagnetic resonance.\citep{Hammel1995oct, Zhang1996apr}

The wide range of MRFM experiments carried out to date have employed distinct polarization schemes: statistical polarization,\citep{Mamin2003nov, Rugar2004jul} Boltzmann polarization,\citep{Garner2004jun, Wagenaar2016jul, DeWit2019feb} and dynamic nuclear polarization,\citep{Chen2013jul, Isaac2016sep, Tabatabaei2024aug}; different spin modulation methods: saturation,\citep{Moore2009dec} adiabatic rapid passage,\cite{Longenecker2012nov} or applying hard NMR pulses;\citep{Verhagen2002feb, Joss2011sep, Alexson2013oct} different detection approaches: forces \cite{Thurber2002mar} and force gradients;\citep{Garner2004jun, Moore2009dec, Boucher2023jan} and distinct detection devices: single-crystal cantilevers,\citep{Longenecker2012nov} nanowires, \citep{Nichol2012feb} and membrane resonators.\citep{Fischer2019Apr} 
The experiments are inherently complex and, pushed to their resolution limits, can have a low signal-to-noise ratio, often requiring minutes,\citep{Degen2009feb, Longenecker2012nov} or even hours,\citep{Rugar2004jul} of signal averaging per data point.
Simulation is required to design an experiment,  discover a first signal, analyze an experiment's signal-to-noise ratio, determine sensitivity and resolution, and is the starting point for spin-density reconstruction in imaging experiments.

Conventionally, we and others have performed MRFM signal simulations in a one-off, proof-of-concept manner.\citep{Nichol2013sep, Mamin2007may, Rugar2004jul, Degen2009feb, Rose2018feb, Thurber2002mar, Moore2009dec, Longenecker2012nov} 
The initial goal was to generate code designed to simulate a single experiment. 
This practice quickly created challenges as experimental plans evolved.
Code was propagated forward from experiment to experiment until an algorithm change triggered a rewrite of all previous code.
It became difficult to optimize and scale up the code. 
The experiments' complexity required collaborators to fully understand the code to make parameter alterations and peer-review the simulations; however, code from the one-off approach was typically understood only by the author, which frustrated collaboration. 
The one-off approach lacked code review and unit tests in a graduate research environment, resulting in erroneous signal estimates. 
For example, more careful coding has revealed that the algorithm used to identify resonant spins in the electron-spin resonance simulation of Fig.~7 in Ref.~\citenum{Moore2009dec} was highly dependent on the grid size. This incorrect algorithm failed to include all of the resonant spins but was paired with a model that overestimated spin saturation. The resulting fit matched experimental data, inadvertently confirming an incorrect signal model.\citep{Isaac2018feb}

Comparing the signal to more rigorous simulations led to a reexamination of the equations governing spin evolution in the experiment; new equations were derived in Ref.~\citenum{Boucher2023sep} that more accurately account for $T_2$ and adiabatic losses during the fast sweeps through resonance experienced by an electron spin beneath a moving cantilever in the MRFM experiment.

In this paper, we present a new open-source Python simulation package for MRFM experiments, \textit{mrfmsim}, designed for modular and extendable development, fast simulation, simple unit tests, and easy collaboration.
The package is built on top of \textit{mmodel},\cite{Sun2023jul} a Python framework that facilitates modular model creation. 
The \textit{mrfmsim} package makes it easy to use community-generated plugins, including experiments and ancillary functionality such as plotting, a command-line interface, and customizations. 
The package shortens the experiment development cycle and facilitates rapid experiment design, simulation, and experimental validation. 
In Section~\ref{section:mrfmsim-architecture}, we present the architecture and workflow of \textit{mrfmsim}. 
We show the challenges of designing a simulation framework for a scientific technique with ongoing development. 
In Section~\ref{section:mrfmsim-implementation}, we show two different experiment examples to simulate and analyze experimental data using \textit{mrfmsim}.
We compare the data analysis against previously published analyses.\citep{Longenecker2012nov, Moore2009dec}
 \section{Architecture}\label{section:mrfmsim-architecture}

The complexity of MRFM experiments and their constantly evolving nature present a unique challenge for experimental simulation. 
First, the experiments share standard components, including a cantilever, a radio frequency (RF) or microwave (MW) source, a magnet, and a sample. 
Yet, they vary in component parameters, detection algorithms, and signal type. 
Therefore, the simulation needs a model-building workflow that can easily reuse shared components while maintaining the flexibility to implement different algorithms and updates. 
Second, collaboration within the research group and with external collaborators is crucial for a scientific technique that is constantly evolving. 
Therefore, the code must be extendable for additional functionalities, readable for communication, and testable for peer review. 
Third, the simulation is used by graduate students with varying programming backgrounds. 
Hence, the package should have a shallow learning curve for executing and modifying a simulation. 
In summary, we want the MRFM simulation to be modular, extendable, readable, and easily testable.

\begin{figure}[t]
\includegraphics[width=\columnwidth]{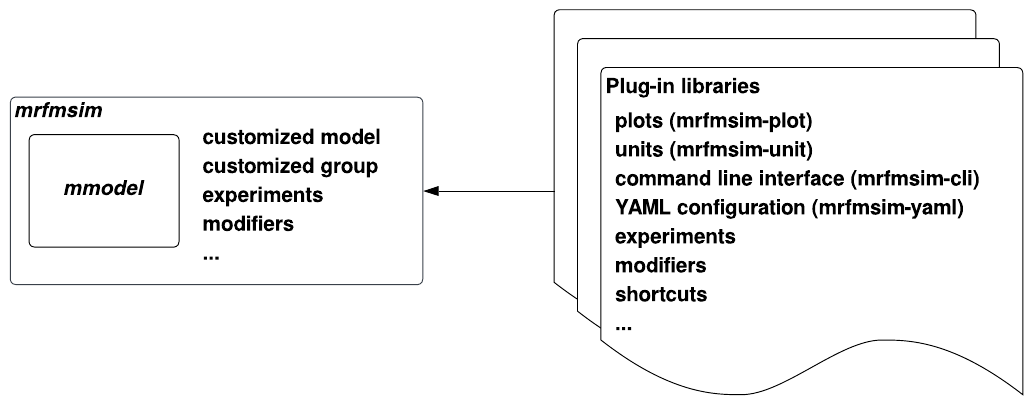}
\caption{\textit{mrfmsim} architecture. The \textit{mrfmsim} package uses \textit{mmodel} as its backend and adds MRFM-specific functionality, including configuration files, modifiers, and shortcuts. Additional functionalities such as experiments, units, plots, and a command-line interface can be added to \textit{mrfmsim} through plug-in libraries.}
    \label{fig:mrfmsim-arch}
\end{figure}

The \textit{mrfmsim} architecture is sketched in Fig.~\ref{fig:mrfmsim-arch}. 
For modularity and readability, we first developed the \textit{mmodel} package for scientific modeling.\citep{Sun2023jul}
The \textit{mmodel} package uses directed acyclic graphs through \textit{NetworkX} \citep{Hagberg2008} to model each experiment, with nodes representing functional steps and edges representing data flows. 
Unlike popular DAG-powered workflow packages such as Dask \citep{Rocklin2015} and Airflow\citep{Apahce2023Jun}, \textit{mmodel} is more lightweight and allows for greater flexibility in modifying the model post-definition. 
For example, MRFM experiments often require simulating the spectra of signals by looping the external magnetic field or the microwave frequency. 
Due to the high computational cost of the underlying matrix calculations, we would like to optimize the signal-simulation code to loop only through functions that involve the field or frequency parameter. 
With the \textit{mmodel} backend, we can quickly generate these loops without duplicating the code or changing the internal model. 
Such modifications to the experiment models are achieved through ``modifiers'' and ``shortcuts'', which add functionalities to the graph nodes and experiment models, respectively. 
DAG-based modeling also improves the readability of the experimental algorithm by plotting a detailed graph that shows the flow of information. 
Finally, \textit{mrfmsim} adds experiment collections, which group similar experiments together and can utilize YAML configuration files to create experiments through plugins.

For extensibility, \textit{mrfmsim} allows users to add additional functionalities through plugin libraries.
Currently, we have defined plugins for defining experiments through configuration files (\textit{mrfmsim-yaml}), three-dimensional plotting (\textit{mrfmsim-plot}), adding units (\textit{mrfmsim-unit}), and adding a command-line interface (\textit{mrfmsim-cli}). 
With this implementation, each research group in the community can develop separate experimental simulation packages to add to the \textit{mrfmsim} package, making the code base lightweight, fast, scalable, and easy to maintain.

\begin{figure}[t]
    \centering\includegraphics[width=\columnwidth]{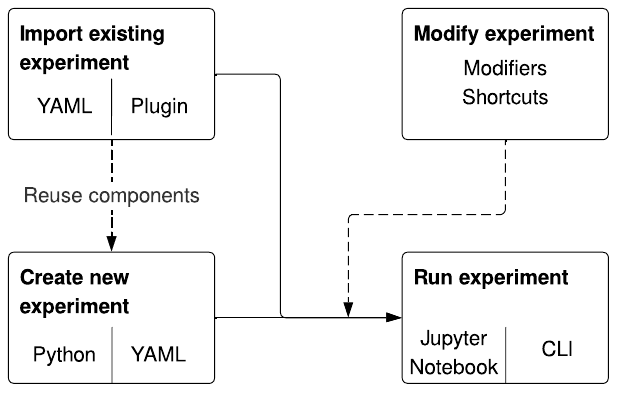}
\caption{\textit{mrfmsim} workflow. A new experiment model can be created using Python scripts or a YAML configuration file (\textit{mrfmsim-yaml}) and converted to a Python object. A new experiment model can reuse components from existing experiments, imported directly from a YAML file or a plugin library. Experiments can be modified using modifiers and shortcuts. Experiment objects can run as a Python script, in a Jupyter notebook, or in the command-line interface (CLI) via the \textit{mrfmsim-cli} plugin.}
    \label{fig:mrfmsim-workflow}
\end{figure}

The workflow for a simulation has four parts --- import, create, modify, and run --- shown in Fig.~\ref{fig:mrfmsim-workflow}. 
In a graduate research setting, typically, there are three types of interactions with the simulation package. 
For existing experiments, the user can directly import the experiment model and execute the model in a way similar to running a regular Python function. 
The experiment is defined in a YAML configuration file or a Python script. 
In this scenario, the user does not need full knowledge of the experiment model's internal code. 
To run the experiment with simple modifications, the user can add modifiers or shortcuts to the experiment, such as looping over a parameter or plotting an intermediate value. 
In this case, the user can inspect the experiment through the model metadata or graph to obtain basic knowledge of the internal model. 
However, no internal code needs to be modified. 

Suppose instead that the user wants to create a new experiment. 
In that case, the user is required to have full knowledge of the DAG modeling capability and understand the internal functions of other models to reuse components. 
The most common use of \textit{mrfmsim}, especially in collaboration with other researchers, is for the first two scenarios, which require minimal code manipulation.
We believe the workflow's clear structure enables a shallow learning curve.

 \section{Case Studies}\label{section:mrfmsim-implementation}

This section examines the spin noise experiment from Ref.~\citenum{Longenecker2012nov} and the CERMIT force gradient experiment from Ref.~\citenum{Moore2009dec}.
We discuss the numerical algorithm for the experiments, the signal lineshape, and analyze experimental data using \textit{mrfmsim}.

\subsection{Spin noise detection with cyclic adiabatic inversions}

The experiment demonstrated in Ref.~\citenum{Longenecker2012nov}, sketched in Fig.~\ref{fig:Longenecker-setup}, utilizes spin noise as the signal for detecting nuclear spins. 
For a random ensemble of spin-1/2 nuclei with a small mean polarization, the variance of the net spin polarization $\Delta N$ is $\sigma^2_{\Delta N} = N$, where $N$ is the total number of spins. 
The standard deviation of such statistical polarization far exceeds the Boltzmann polarization in a small detection volume. 
The signal variance or ``spin noise'' can be used as the MRFM signal.\citep{Mamin2003nov, Rugar2004jul, Longenecker2012nov} 
A proton spin in the setup experiences a resonance offset $\Delta B_0$,
\begin{equation}
    \Delta B_0 = B\st{ext} + B^\mathrm{tip}_z - 2\pi \frac{f_\mathrm{rf}}{\gamma_\mathrm{p}},
\end{equation}
where $\gamma_\mathrm{p} = 2 \pi \times \SI{42.577}{\mega\hertz\per\tesla}$ is the proton gyromagnetic ratio, 
$B\st{ext}$ is an external magnetic field, and 
$B^\mathrm{tip}_z$ is the tip field. 
The force variance signal per spin   $\sigma^2_\mathrm{spin}$ at resonance offset $\Delta B_0$ can be calculated as \citep{Mamin2003nov, Rugar2004jul, Longenecker2012nov}
\begin{equation}\label{eq:res-psf}
    \sigma^2_\mathrm{spin} = A \mu_\mathrm{p}^2\left(\frac{\partial B_z^\mathrm{tip}}{\partial x}\right)^2\eta (\Delta B_0),
\end{equation}
where $A$ is a constant scaling factor, 
$\mu_\mathrm{p} = \SI{1.4e26}{\joule\per\tesla}$ is the proton magnetic moment, 
$\eta (\Delta B_0)$ is a function that characterizes the off-resonance spin response.

For a cyclic inversion protocol with triangle-wave frequency modulation, the off-resonance response is thought to be well approximated by \citep{Mamin2003nov, Rugar2004jul, Longenecker2012nov}
\begin{equation}\label{eq:eta(dB0)}
\eta (\Delta B_0)=
\begin{cases} 
      \cos^2{\left(\dfrac{\gamma_\mathrm{p}\Delta B_0}{2\Delta f_\mathrm{FM}}\right)} & \mathrm{for}\; \Delta B_0 \leq \pi \Delta f_\mathrm{FM}/\gamma_\mathrm{p}\\
      0 & \mathrm{otherwise}, \\
   \end{cases}
\end{equation}
where $\Delta f_\mathrm{FM}$ is the peak-to-peak frequency deviation.
The total variance signal is the sum of the force variance of all the spins in the detection volume.

\subsubsection{Magnetic field from a rectangular magnet}
The magnetization $\bm{M}$ for a rectangular magnet polarized along the $z$-direction is
\begin{equation}
\bm{M} = \mu_0 M_\mathrm{s} \bm{u}_z,
\end{equation}
where $\mu_0M_\mathrm{s}$ is the saturation magnetization of the magnet, and $\bm{u}_z$ is a unit vector in the $z$-direction. 
For the rectangular magnet of interest, we define the dimensions of the magnet to span from $x_1$ to $x_2$ in the $x$-direction, from $y_1$ to $y_2$ in the $y$-direction, and from $z_1$ to $z_2$ in the $z$-direction. We model the magnet as a series of coils, where each loop is replaced by a layer of continuous current density. The magnetic potential,l summing over each face of the magnet, is
\begin{multline}
\Phi(x,y,z) = \dfrac{\mu_0 M_\mathrm{s}}{4\pi} \sum_{i=1}^2 \sum_{j=1}^2 \sum_{k=1}^2 (-1)^{(i+j+k)}\\
\biggl(y_i + (z-z_k)\arctan \left(\dfrac{y - y_i}{z-z_k}\right)
+ (x-x_i)\log(y-y_j + R) \\
- (z-z_k)\arctan \left( \dfrac{(x-x_i)(y-y_j)}{(z-z_k) R} \right) \\
+ (y-y_j)\log (x-x_i + R)\biggr),
\label{Eq:fullPhi}
\end{multline}
with $R = \sqrt{(x-x_i)^2 + (y-y_j)^2 + (z-z_k)^2}$.
Given that the z-component of the magnetic field $B_z$ is
\begin{equation}
B_z (x,y,z) = -{\nabla}(\Phi(x,y,z))\cdot\bm{u}_z.
\end{equation}
we can obtain the field component $B_z$ with Eq.~\ref{Eq:fullPhi},
\begin{multline}
B_z(x,y,z) = \dfrac{\mu_0 M_\mathrm{s}}{4\pi}
\sum_{i=1}^2 \sum_{j=1}^2 \sum_{k=1}^2 (-1)^{(i+j+k)}\\
\arctan\left(\dfrac{(x-x_i)(y-y_j)}{(z-z_k) R}\right).
\end{multline}
The result is consistent with the field of permanent rectangular magnetic presented in Ref.~\citenum{Yang1990dec}.
The first derivative of the $B_z$ field in the $x-$direction is,
\begin{equation}
\dfrac{\partial B_z}{\partial x} = \dfrac{\mu_0 M_\mathrm{s}}{4\pi} \sum_{i=1}^2 \sum_{j=1}^2 \sum_{k=1}^2 (-1)^{(i+j+k)}
 \dfrac{(y-y_j)(z-z_k)}{\left((x-x_i)^2 +(z-z_k)^2\right)R}.
\end{equation}
The second derivative of the magnetic field is then given by
\begin{multline}
    \frac{\partial^2 B_z}{\partial x^2} = \frac{\mu_0 M_\mathrm{s}}{4\pi} \sum_{i=1}^2 \sum_{j=1}^2 \sum_{k=1}^2 (-1)^{(i+j+k)} \\
    \frac{-(x-x_i)(y-y_j)(z-z_k)}
         {\left((x-x_i)^2 +(y-y_j)^2+(z-z_k)^2\right)^{3/2} \left((x-x_i)^2 + (z-z_k)^2\right)^2} \\
    \times \left(3(x-x_i)^2 + 2(y-y_j)^2 + 3(z-z_k)^2\right).
\end{multline}

\subsubsection{Comparison with experimental data}

\begin{figure}[t]
    \centering
\includegraphics[width=\columnwidth]{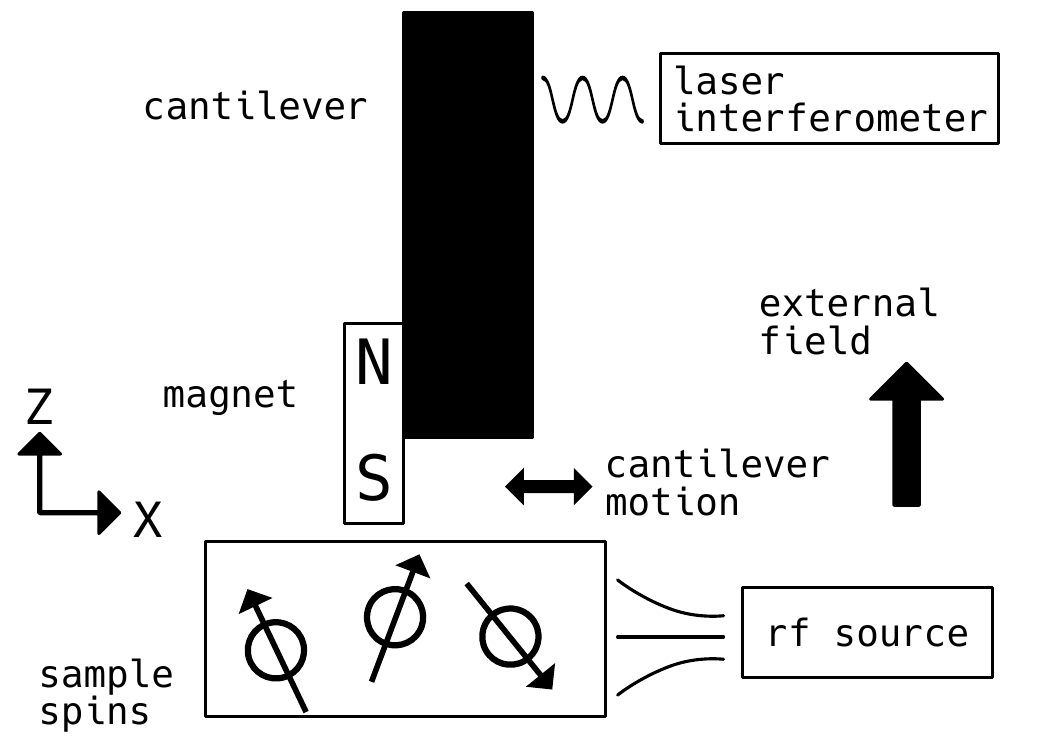}
\caption{Schematic for Ref.~\citenum{Longenecker2012nov} spin-noise experiment. \textbf{Cantilever:} Silicon cantilever with the dimensions of 
    \SI{4}{\micro\meter} $\times$ \SI{200}{\micro\meter} $\times$ \SI{0.4}{\micro\meter}. 
The cantilever has a resonance frequency $f_\mathrm{c}= \SI{6644}{\hertz}$, an intrinsic quality factor of $Q=\num{8.4e4}$ in vacuum, and a spring constant of $k=\SI{1.0}{\milli\newton\per\meter}$. \textbf{Magnet:} cobalt magnet with dimensions \SI{225}{\nano\meter} $\times$ \SI{79}{\nano\meter} $\times$ \SI{1494}{\nano\meter}. The magnet extends past the leading edge of the cantilever by \SI{300}{\nano\meter}. \textbf{RF source:} lithographically defined copper microwire on a silicon substrate. \textbf{Sample:} a \SI{40}{\nano\meter} uniform layer of polystyrene (molecular weight = 200000) film, spin-coated onto the microwire substrate.} 
    \label{fig:Longenecker-setup}
\end{figure}

As shown in Fig.~\ref{fig:Longenecker-setup},
the cantilever was aligned to the $z$ direction along with the external field. 
The cantilever motion was in the $x$ direction. The magnet was attached to the cantilever tip, and the cantilever position was monitored using a laser interferometer. 
Simulation parameters are given in the Fig.~\ref{fig:Longenecker-setup} caption and in Table.~\ref{table:mrfmsim-longenecker-exp-parameters}. 
The saturation magnetization $\mu_0M_\mathrm{sat}$ of the cobalt magnet was estimated to be \SI{1.91 \pm 0.03}{\tesla} using SQUID measurements. 
However, due to the uncertainty in the spring constant, we used a cobalt bulk saturation magnetization of \SI{1.8}{\tesla} for the fit. 
The radio frequency source was a lithographically defined copper microwire on a silicon substrate. 
The sample was a \SI{40}{\nano\meter}-thick uniform polystyrene layer. 

\begin{table}[!htbp]
\centering
\caption{Simulation parameters for the spin noise experiment (Fig.~\ref{fig:Longenecker-setup}) with a polystyrene film sample and a cobalt rectangular prism magnet tipped cantilever.}
\begin{tabular}{lll}
\hline
variable & definition & numerical value \\ 
\hline
$\gamma_\mathrm{p}$ & gyromagnetic ratio & $2\pi \times \SI{42.577}{\mega\hertz\per\tesla}$ \\
$T_2$ & spin-spin relaxation & \SI{5}{\micro\second} \\
$T_1$ & spin-lattice relaxation & \SI{10}{\second} \\
$\rho$ & spin density & \SI{49}{\per\nano\meter\cubed} \\
$B\st{ext}$ & external field & \SI{2630.5}{\milli\tesla} \\
$T$ & temperature & \SI{5.5}{\kelvin} \\
 $\Delta f_\mathrm{FM}$ & peak-to-peak FM deviation & \SI{2}{\mega\hertz} \\
 & magnet dimension & \SI{225}{\nano\meter} $\times$ \SI{79}{\nano\meter} $\times$ \SI{1494}{\nano\meter}\\
$\mu_0M_\mathrm{s}$ & saturation magnetization & \SI{1.8}{\tesla}\\
 & sample grid size & \SI{400}{\nano\meter} $\times$ \SI{400}{\nano\meter} $\times$ \SI{40}{\nano\meter} \\
 & sample grid step size & \SI{1}{\nano\meter} \\
\hline
\end{tabular}

\label{table:mrfmsim-longenecker-exp-parameters}
\end{table}

\begin{figure*}[!htbp]
    \centering
    \includegraphics[width=0.75\textwidth]{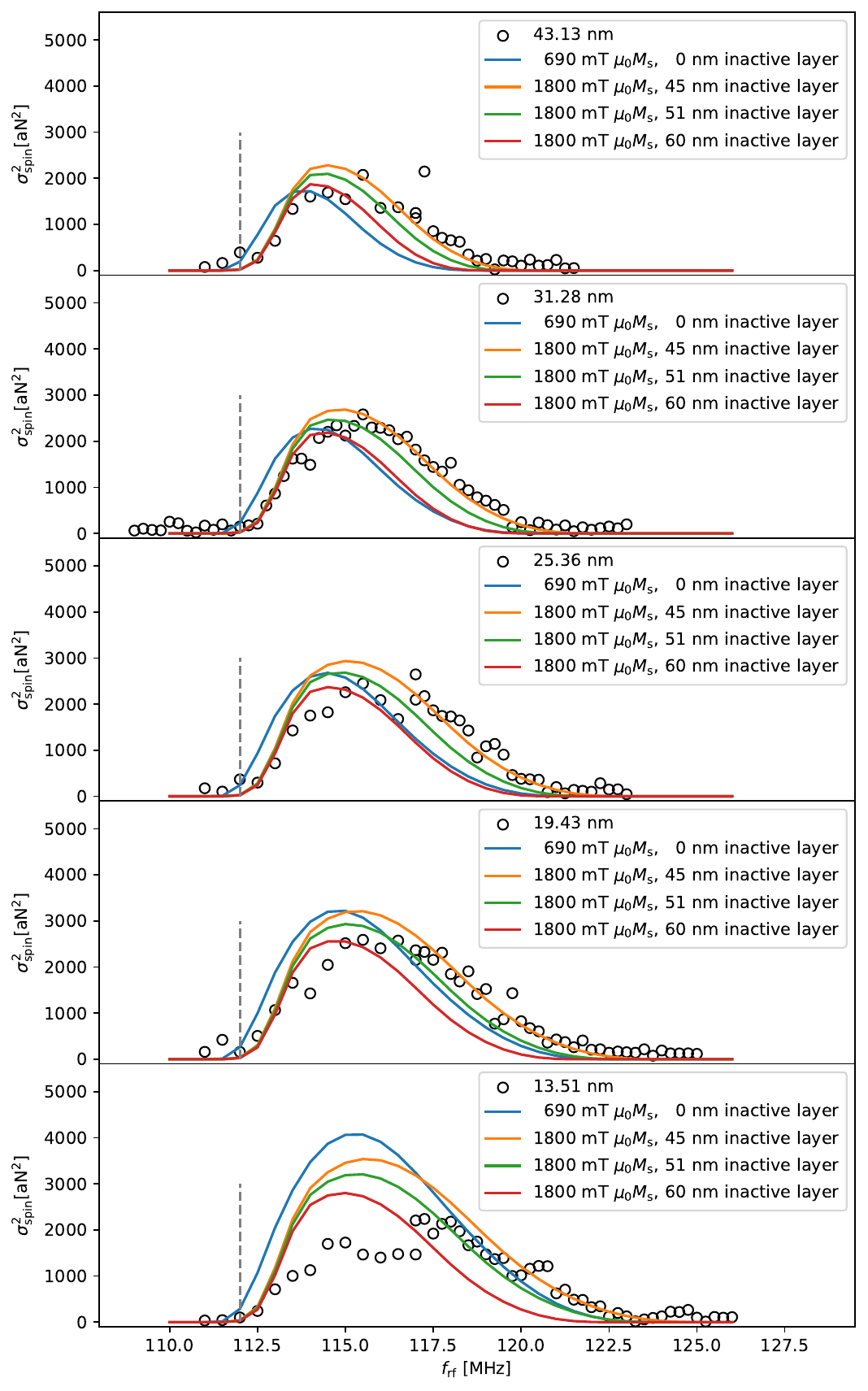}
\caption{Experiment data from Ref.~\citenum{Longenecker2012nov} at different tip-sample separations (circle). Signals at different magnet saturation magnetizations ($\mu_0 M_\mathrm{s}$) with different damage layers are simulated. The sample used in the simulations is a \SI{40}{\nano\meter} thick polystyrene film with a spin density of \SI{49}{\per\nano\meter\cubed}, $T_1$ of \SI{10}{\second}, $T_2$ of \SI{5}{\micro\second}, external field $B_0$ of \SI{2360.5}{\milli\tesla}, and $\Delta f_\mathrm{FM}$ of \SI{2}{\mega\hertz}. The gray dashed line is the Larmor frequency expected in the absence of the tip field.}
    \label{fig:Longenecker-fit}
\end{figure*}

Spin noise variance signal versus radio frequency data were collected at several different tip-sample separations. 
In Ref.~\citenum{Longenecker2012nov}, the authors discussed two different magnet models to explain the data.
The first model was a magnet with a reduced saturation magnetization of \SI{0.69}{\tesla}. 
The second model was a magnet with a \SI{51}{\nano\meter} inactive layer at the leading edge. 
The inactive layer can result from extraneous space, as suggested by Longenecker, or from damage during the fabrication process.
It is important to note that the spectrum at each tip-sample separation was treated with an unreported overall scaling constant $A$ (Eq.~\ref{eq:res-psf}). 
Here, we used $A=1$. 
Using \texttt{mrfmsim.Experiment.IBMCyclic}, we simulated both models presented in the paper and added two additional models with a leading-edge inactive layer of \SI{45}{\nano\meter} and \SI{60}{\nano\meter}.
The data and the simulated results are shown in Fig.~\ref{fig:Longenecker-fit}. 

We observe that the signal onsets at the Larmor frequency (gray dashed line) and offsets at different radio frequencies depending on the tip-sample separation.
Interestingly, we observe that the \SI{45}{\nano\meter} model provides a better fit to the lineshape, particularly at high radio frequencies. 
To further explore the simulation, we analyze the lineshape visually.

\subsubsection{Signal lineshape}\label{subsubsec:Joni-lineshape}

\begin{figure}[t]
    \centering
    \includegraphics[width=\columnwidth]{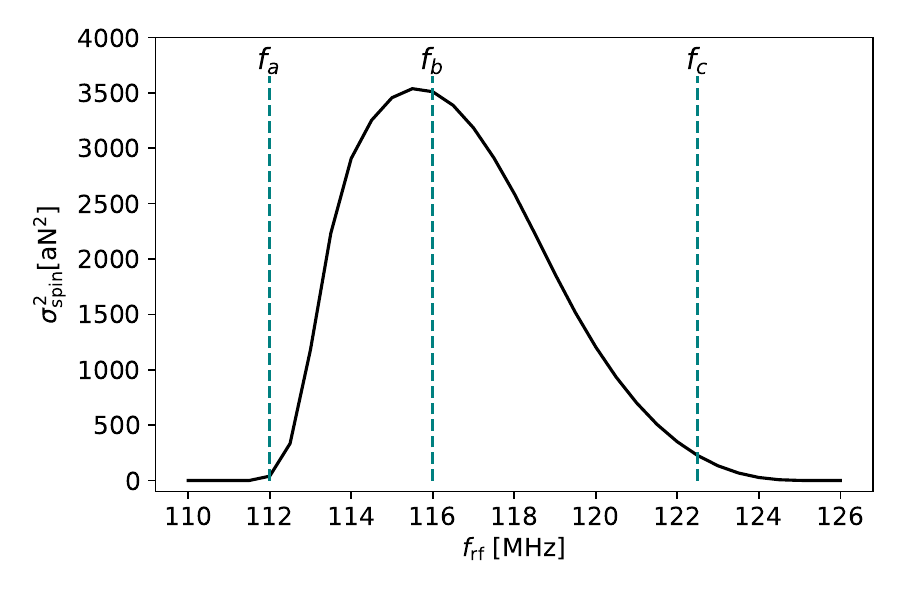}
\caption{Simulated signal of experiment in Ref.~\citenum{Longenecker2012nov} with the magnet saturation magnetization of \SI{1.8}{\tesla}, an inactive layer of \SI{45}{\nano\meter}, and at a tip-sample separation of \SI{13.53}{\nano\meter}. The dashed vertical teal lines at \SI{112}{\mega\hertz}, \SI{116}{\mega\hertz}, and \SI{122.5}{\mega\hertz} indicate the points where the lineshape is visually analyzed.}
    \label{fig:Longenecker-freqs}
\end{figure}

\begin{figure*}[!ht]
\centering
\includegraphics[width=0.9\textwidth]{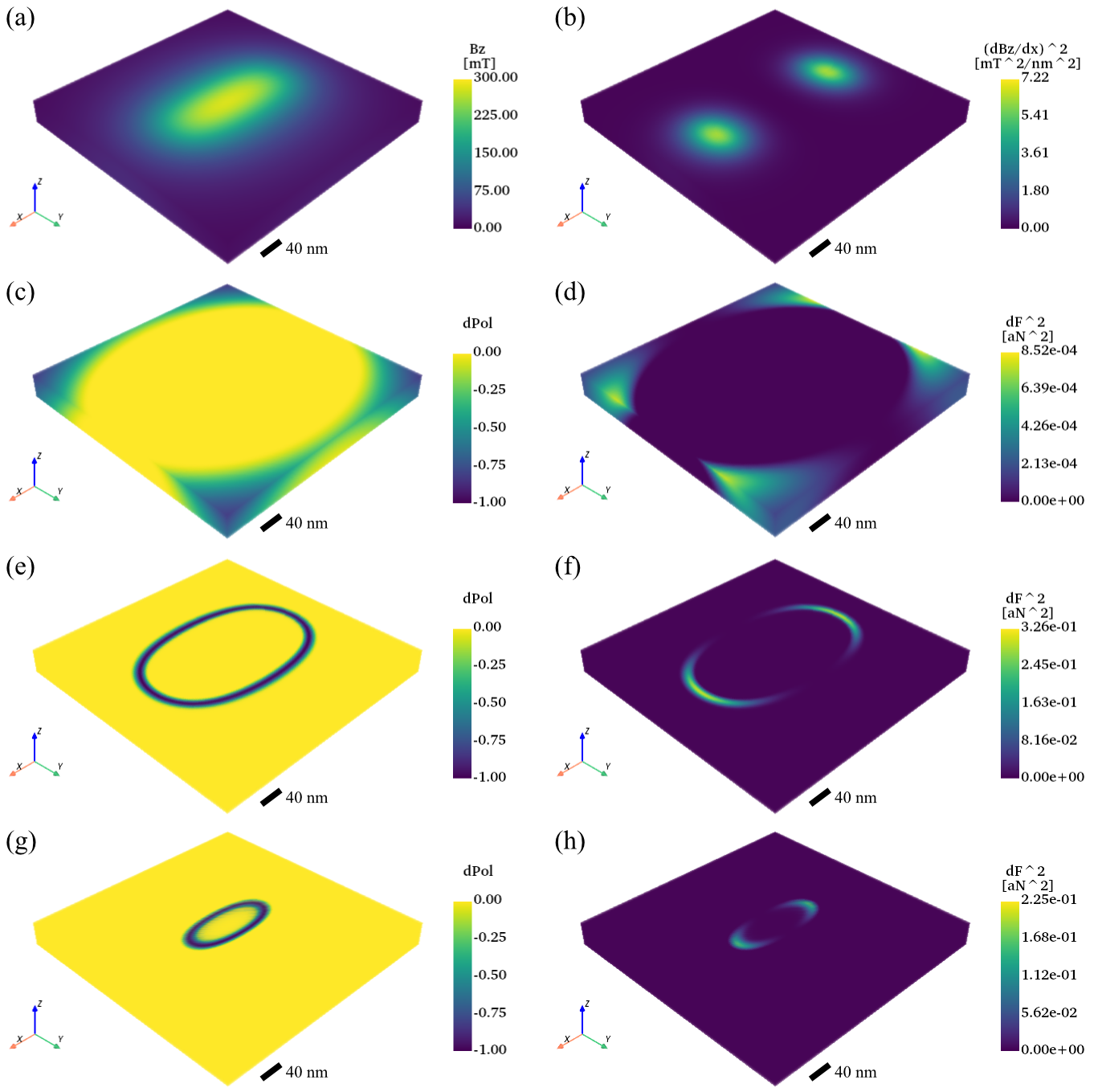}
\caption{Experiment lineshape render for Ref.~\citenum{Longenecker2012nov} at a tip-sample separation of \SI{13.53}{\nano\meter} with the magnet saturation magnetization of \SI{1.8}{\tesla}, and a magnet inactive layer of \SI{45}{\nano\meter}. The sample grid is \SI{400}{\nano\meter} $\times$ \SI{400}{\nano\meter} $\times$ \SI{40}{\nano\meter} with a \SI{1}{\nano\meter} step size.
(a) Tip field $B_\mathrm{tip}$ overlayed with the sample film. 
(b) The square of the tip field gradient $(\partial B_z^\mathrm{tip}/{\partial x})^2$ overlayed with the sample film. The spins in resonance at radio frequencies of \SI{112}{\mega\hertz}, \SI{116}{\mega\hertz}, and \SI{122.5}{\mega\hertz} are shown in (c), (e), and (g), respectively.
The signal distributions (Eq.~\ref{eq:res-psf}) at the three radio frequencies are shown in (d), (f), and (h). 
Volumes are rendered using the \textit{PyVista} package.\citep{Sullivan2019may}}
    \label{fig:Longenecker-lineshape}
\end{figure*}

For a given external field, tip magnetization, applied radio frequency, tip-sample separation, and experimental protocol, spins in and near the resonance condition form a dome-shaped sensitive slice, characterized by Eq.~\ref{eq:eta(dB0)}. The signal distribution for individual grid points of a uniformly distributed spin sample can be determined by Eq.~\ref{eq:res-psf}.
The total sample grid size was increased until the truncation of the line shape was negligible. The sample step size was chosen to be comparable to the sensitive slice width, while keeping the number of sample grid points reasonably small to reduce computational time. 
To analyze the experiment's lineshape, we examined the sensitive slice and signal distribution at three different radio frequencies. The three frequencies correspond to the signal's onset (Larmor frequency), peak, and offset.
Spins in the sensitive slice are inverted and contribute to the signal. 
Figure~\ref{fig:Longenecker-lineshape} shows the rendered sensitive slice and signal distribution result. 
Here, we select the simulated signal at \SI{13.53}{\nano\meter}, and the tip magnet has a \SI{45}{\nano\meter} leading-edge inactive layer. The frequencies were chosen at \SI{112}{\mega\hertz}, \SI{116}{\mega\hertz}, and \SI{122.5}{\mega\hertz}; the simulated signal and frequency positions are shown in Fig.~\ref{fig:Longenecker-freqs}.
Figure \ref{fig:Longenecker-lineshape}(a) shows that the tip magnet field is elliptically shaped due to the asymmetry of the magnet tip.
Figure \ref{fig:Longenecker-lineshape}(b) is the field gradient, $(\partial B_z^\mathrm{tip}/{\partial x})^2$, and we can see that it is concentrated at two sides of the magnet.
At the Larmor frequency, the sensitive slice consists of spins that experience little tip field or tip-field gradient (Fig.~\ref{fig:Longenecker-lineshape}(c)); these spins contribute very little to the signal (Fig.~\ref{fig:Longenecker-lineshape}(d)). 
With increased radio frequency, the sensitive slices become bowl-shaped, and the signal reaches a maximum when it overlaps strongly with the tip gradient (Fig.~\ref{fig:Longenecker-lineshape}(e-f)). Finally, at the highest radio frequency, due to the tip-sample separation, the sensitive slice no longer passes through the sample, and the signal decreases (Fig.~\ref{fig:Longenecker-lineshape}(g-h)).

In conclusion, mrfmsim correctly calculates both the size of the Ref.~\citenum{Longenecker2012nov} spin-noise signal and its lineshape.  
The lineshape can be understood by plotting the sensitive slices as a function of radio frequency. 
There is no signal when the radio frequency is smaller than the Larmor frequency because no spins meet the resonance condition, regardless of the tip-sample separation. 
The signal roll-off at higher radio frequencies is due to the decreasing overlap between the sensitive slice and the sample. Therefore, as the tip-sample separation increases, the high-frequency end of the lineshape shifts to a lower radio frequency. 
This observation matches both the data and Fig.~\ref{fig:Longenecker-fit}.
The constant scaling factor $A$ in Eq.~\ref{eq:res-psf} was used by previous authors to adjust the signal amplitude without altering the lineshape. This could be used to adjust for errors, for example, in the cantilever spring constant.
In this section's simulation, we set $A=1$ and reproduced the signal lineshape and, remarkably, the absolute signal size.
That Eq.~\ref{eq:res-psf} correctly predicts the signal size is an important independent check of its validity.
 \subsection{Force gradient detection with the CERMIT protocol}\label{subsection: mrfmsim-moore-tdo}

\begin{figure}[t]
    \centering
\includegraphics[width=\columnwidth]{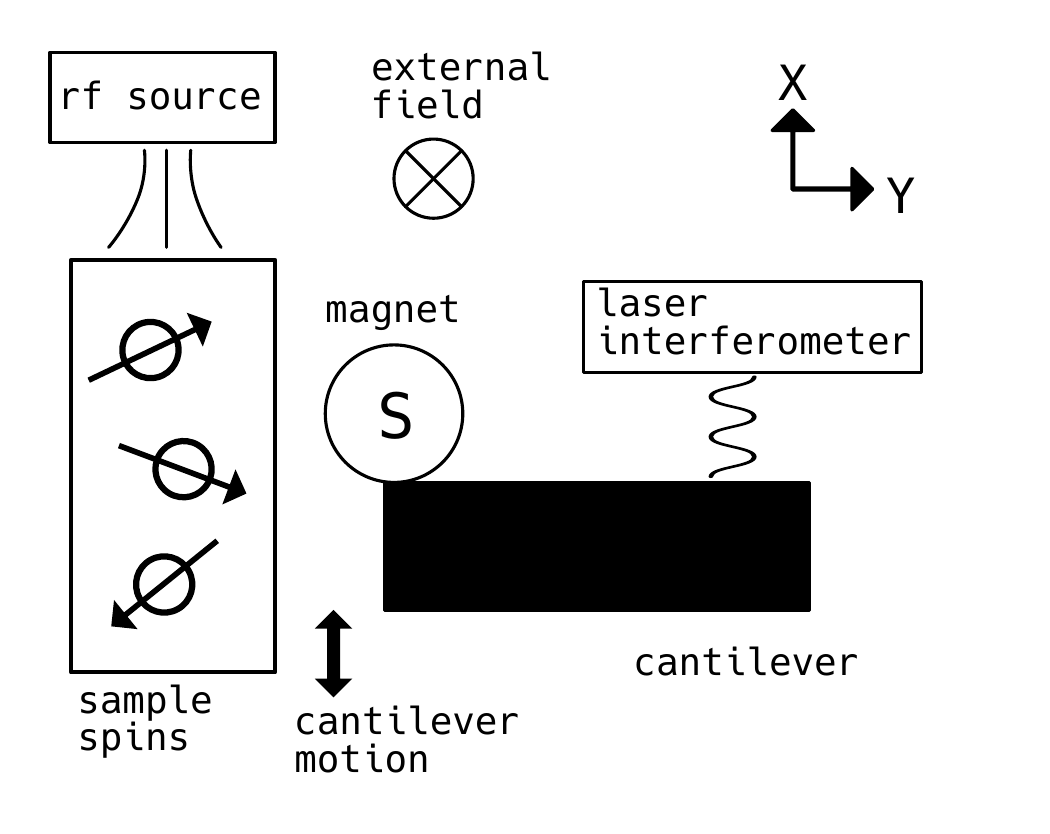}
\caption{Experiment schematic for Ref.~\citenum{Moore2009dec}.~\textbf{Cantilever:} resonance frequency $f_\mathrm{c} = \SI{5512}{\hertz}$ and a spring constant of $k=\SI{0.78}{\milli\newton\per\meter}$. \textbf{Magnet:} nickel magnet with a diameter of \SI{2752}{\nano\meter}.  \textbf{RF source:} coplanar waveguid. \textbf{Sample:} a \SI{200}{\nano\meter} uniform layer of tempamine-doped perdeuterated polystyrene.} 
    \label{fig:Moore-setup}
\end{figure}

Reference \citenum{Moore2009dec} presented a modulated CERMIT (Cantilever-Enabled Readout of Magnetization Inversion Transients) protocol to detect electron spin resonance from nitroxide spin labels using cantilever-synchronized microwave pulses. The Ref.~\citenum{Moore2009dec} experiment is sketched in Fig.~\ref{fig:Moore-setup}.
Reference \citenum{Boucher2023sep} introduced a signal model accounting for both adiabatic losses and $T_2$ relaxation arising from cantilever motion during the MW pulses.
That paper's signal model is valid in the limit of $T_2 \ll T_1$, where the transverse magnetization reaches a pseudo-equilibrium with the slowly evolving longitudinal magnetization.
We define a unitless time variable $\tau$ and relaxation time variables $\alpha$ and $\beta$ as
\begin{equation}
    \tau = \gamma_\mathrm{e} B_1 t,
\end{equation}
\begin{equation}
    \alpha = \frac{1}{\gamma_\mathrm{e} B_1 T_1},
\end{equation}
and
\begin{equation}
    \beta = \frac{1}{\gamma_\mathrm{e} B_1 T_2},
\end{equation}
where $\gamma_\mathrm{e}$ is the electron gyromagnetic ratio, and $B_1$ is the irradiation intensity in the rotating frame. The experimental parameters for Ref.~\citenum{Moore2009dec} are shown in Table.~\ref{table:moore-exp-parameters}. 
The cantilever period is \SI{0.2}{\milli\second}, the $T_2$ for the sample is \SI{0.45}{\micro\second}, and $T_1$ is \SI{1.3}{\milli\second}.
Therefore, most spins in the regime of interest spend less than $T_1$ within the homogeneous linewidth, and the experimental $B_1$ of \SI{2.89}{\micro\tesla} is well below the critical field of \SI{6.3}{\milli\tesla} calculated using Eq.~14 in Ref.~\citenum{Boucher2023sep}.
In this low-$B_1$ regime, we can determine numerically the final magnetization $M_z^\mathrm{final}$ during a cantilever sweep from $\tau_i$ to $\tau_f$, and sweep time $\tau_f - \tau_i \ll T_1$ as follows,
\begin{equation}
M_z^\mathrm{final} 
  \approx e^{-R(\tau_i, \tau_f)} M_z^\mathrm{initial},
  \label{eq:Mz(tau0)}
\end{equation}
and
\begin{equation}
\label{eq:R(tau_f)}
R(\tau_i, \tau_f) 
  = \frac{\arctan{(\pi \alpha_1 \tau_f / \beta)} - \arctan{(\pi \alpha_1 \tau_i / \beta)}}{\pi \alpha_1},
\end{equation}
where
\begin{equation}
    \alpha_1 = \frac{1}{\pi \gamma_\mathrm{e} B_1^2}\frac{d\Delta B_0}{dt}
\end{equation}
is a unitless sweep-rate parameter, assuming the resonance offset changes linearly in time. The resonance offset $\Delta B_0$ is calculated as $\Delta B_0 = B\st{ext} + B_z\sr{tip} - 2\pi f\st{mw}/\gamma_\mathrm{e}$, where $B_z\sr{tip}$ is the tip magnetic field, $f\st{mw}$ is the applied MW frequency, and $\gamma_\mathrm{e}$ is the electron gyromagnetic ratio.

At time $\tau = 0$, the system is at the resonance condition. The argument of the arctan function can be written as
\begin{equation}
\frac{\pi \alpha_1 \tau}{\beta} 
  = \gamma_\mathrm{e} T_2 \frac{d \Delta B_0}{d t} t = \gamma_\mathrm{e} T_2 \Delta B_0 (t).
\label{eq:a/b}
\end{equation}
Additionally, since we approximate that the cantilever velocity $v\sr{tip}$ remains the same during the sweep, we can calculate the change in the magnetic field offset, 
\begin{equation}
\frac{d \Delta B_0}{d t} = \frac{\partial \Delta B_0}{\partial x} \frac{\partial x}{\partial t} = \frac{\partial \Delta B_0}{\partial x} v\sr{tip}.
\end{equation}
Finally,
\begin{equation}\label{eq:mzf/mzi}
\frac{M_z\sr{final}}{M_z\sr{initial}} \approx e^{-R(\tau_i, \tau_f)},
\end{equation}
where 
\begin{multline}\label{eq:R-ratio}
R(\tau_i, \tau_f) = \frac{\gamma_\mathrm{e}  B_1^2}{B_{zx}\sr{tip}v\sr{tip}}\\ (\arctan{(\gamma_\mathrm{e} T_2 \Delta B_0(\tau_f)}) - \arctan{(\gamma_\mathrm{e} T_2 \Delta B_0(\tau_i))}),
\end{multline}
and $B_{zx}\sr{tip} = \partial \Delta B_0/\partial x$ is the the field gradient of $B_z\sr{tip}$ in the $x$ direction.

\begin{table}[t]
\centering
\caption{Simulation parameters for the CERMIT experiment (Fig.~\ref{fig:Moore-setup}) with a tempamine-doped perdeuterated polystyrene film sample and a nickel spherical magnet tipped cantilever.}
\begin{tabular}{lll}
\hline
variable & definition & numerical value \\ 
\hline
$\gamma_\mathrm{e}$ & electron gyromagnetic ratio & $2\pi \times \SI{28.0}{\giga\hertz\per\tesla}$ \\
$T_2$ & spin-spin relaxation & \SI{0.45}{\micro\second} \\
$T_1$ & spin-lattice relaxation & \SI{1.3}{\milli\second} \\
$B_1\sr{crit}$ & $(1/T_2 - 1/T_1)/(2 \gamma_\mathrm{e} )$ & \SI{6.3}{\milli\tesla}  \\ 
$B_1$ & transverse field & \SI{2.89}{\micro\tesla} \\
$B\st{homog}$ & $1/(\gamma_\mathrm{e} \, T_2)$ & \SI{12.6}{\milli\tesla} \\
$B\st{sat}$ & $1/(\gamma_\mathrm{e} \sqrt{T_1 T_2})$ & \SI{0.24}{\milli\tesla} \\ 
$f_\mathrm{c}$ & cantilever frequency & \SI{5512}{\hertz} \\
$k_\mathrm{c}$ & cantilever spring constant & \SI{0.78}{\milli\newton\per\meter} \\
$x_\mathrm{pk}$ & cantilever amplitude & \SI{164}{\nano\meter} \\
$f_\mathrm{mw}$ & microwave frequency & \SI{18.1}{\giga\hertz} \\
$\rho$ & spin density & \SI{0.0241}{\per\nano\meter\cubed}\\
$T$ & temperature & \SI{11}{\kelvin} \\
$M_s$ & saturation magnetization & \SI{523}{\milli\tesla} \\
$r$ & magnet radius & \SI{1376}{\nano\meter}\\
& sample grid size & \SI{10}{\micro\meter} $\times$ \SI{0.2}{\micro\meter} $\times$ \SI{22}{\micro\meter} \\
& sample grid step size & \SI{20}{\nano\meter} \\

\hline
\end{tabular}

\label{table:moore-exp-parameters}
\end{table}

The above derivation accounts for the magnetization change after a single pulse.
For the multi-pulse CERMIT experiment, MW pulses are applied every $n$ cantilever cycles, typically $n=$ 1 to 3. We need to account for $T_1$ relaxation of longitudinal magnetization between pulses.
Assuming that the system is at steady-state, we use the simplified notation that before and after the pulse $p$, the magnetization is $M^{-} = M_z^\mathrm{eq}(\tau_p^-)$ and $M^{+} = M_z^\mathrm{eq}(\tau_p^+)$, respectively; the magnetization ratio $r$ is therefore $r = M^{+}/M^{-} =e^{-R(\tau_p^+, \tau_p^-)}$.
The magnetization relaxes towards the initial magnetization $M_0$,
\begin{equation}
    M^{-} - M_0 = (M^{+}- M_0)e^{-t \big/ T_1},
\end{equation}
where $t$ is the time between pulses.
The microwave pulse occurs simultaneously during each cantilever cycle, and $R$ remains constant regardless of the system's magnetization state. 
With the pulse interval $\Delta t$, the change in magnetization before the relaxation is
\begin{equation}
    \Delta M^{+} = M^{+} - M_0 = \frac{r - 1}{1 - r e^{-\Delta t\big/T_1}} M_0.
\end{equation}
Therefore the average change in magnetization $\Delta M_z^\mathrm{avg}$ is
\begin{equation}\label{eq:dMz_avg}
    \Delta M_z^\mathrm{avg} =  \frac{\int_{0}^{\Delta t}{\Delta M^{+}e^{-t\big/T_1}} \,dt}{\Delta t} = \frac{(r - 1) (1 - e^{-t\big/T_1})}{1-re^{-t\big/T_1}}\frac{T_1}{\Delta t} M_0.
\end{equation}
Given that the cantilever motion is small compared to the magnet radius, the force-gradient signal resulting from the magnetization change summed over $\bm{r}_j$ is
\begin{equation}\label{eq:moore-final}
    \delta f = -\
\frac{\sqrt{2}f_\mathrm{c}}{2\pi k_\mathrm{c}}\sum_j \Delta M_z^\mathrm{avg}(\bm{r}_j) \frac{\partial ^2 B_z ^\mathrm{tip} (\bm{r}_j)}{\partial x^2},
\end{equation}
where $f_\mathrm{c}$ is the cantilever frequency, $k_\mathrm{c}$ is the cantilever spring constant, and the $\sqrt{2}/\pi$ factor accounts for the conversion from a constant frequency shift to a root-mean-square lock-in output.

The Eq.~(\ref{eq:mzf/mzi}), (\ref{eq:R-ratio}), (\ref{eq:dMz_avg}), and (\ref{eq:moore-final}) are implemented in the \texttt{mrfmsim.experiment.CermitTD} experiment class. These equations assume that the spin relaxation $T_1 \ll T_2$, and the resonance offset changes linearly in time for the spins during the cantilever sweep.

\subsubsection{Magnet field from a spherical magnet}
The magnetic field at a point ($x, y, z$) from a magnet centered at ($x_0, y_0, z_0$) with radius $r$,\citep{Griffiths2017}
\begin{equation}
B_z = \dfrac{\mu_0 M_s}{3} \left(3 \dfrac{Z^2}{R^5} - \dfrac{1}{R^3} \right),
\label{Eq:sphere_Bz}
\end{equation}
where $\mu_0 M_s$ is the magnet’s magnetization and $R$ is given by 
\begin{equation}
R = \sqrt{X^2 + Y^2 +Z^2},
\end{equation}
and $X= (x - x_0)/r$, $Y = (y - y_0)/r$, and $Z = (z - z_0)/r$ are the normalized distances to the center of the magnet. The first and second derivatives of the magnetic field with respect to $x$ are,
\begin{equation}
    \dfrac{\partial B_z}{\partial x} = \dfrac{\mu_0 M_s}{r} X \left(\dfrac{1}{R^5} - 5 \dfrac{Z^2}{R^7} \right),
\label{Eq:sphere_Bzx}
\end{equation}
and 
\begin{equation}
\dfrac{\partial^2 B_z}{\partial x^2} = \dfrac{\mu_0 M_s}{r^2}
\left(\dfrac{1}{R^5} - 5 \dfrac{X^2}{R^7} - 5 \dfrac{Z^2}{R^7} + 35 \dfrac{X^2 Z^2}{R^9} \right).
\label{Eq:sphere_Bzxx}
\end{equation}

\subsubsection{Experimental data}

Compare the setup in Fig.~\ref{fig:Moore-setup} to Fig.~\ref{fig:Longenecker-setup}, the cantilever's long axis was aligned along the $y$ direction, while the external field was aligned along the $z$ direction.
The magnet was attached to the cantilever, and the cantilever oscillated in the $x$ direction. 
The experiment parameters are given in Ref.~\ref{table:moore-exp-parameters}.
We want to point out that the data used in this paper was acquired during the same run of the Ref.~\citenum{Moore2009dec} experiment but at a different sample location. 
The tip magnet was a spherical nickel magnet with a radius of \SI{1376}{\nano\meter}. 
The sample was a \SI{200}{\nano\meter} thick tempamine-doped perdeuterated polystyrene film.

\begin{figure*}[!htbp]
    \centering
\includegraphics[width=0.65\textwidth]{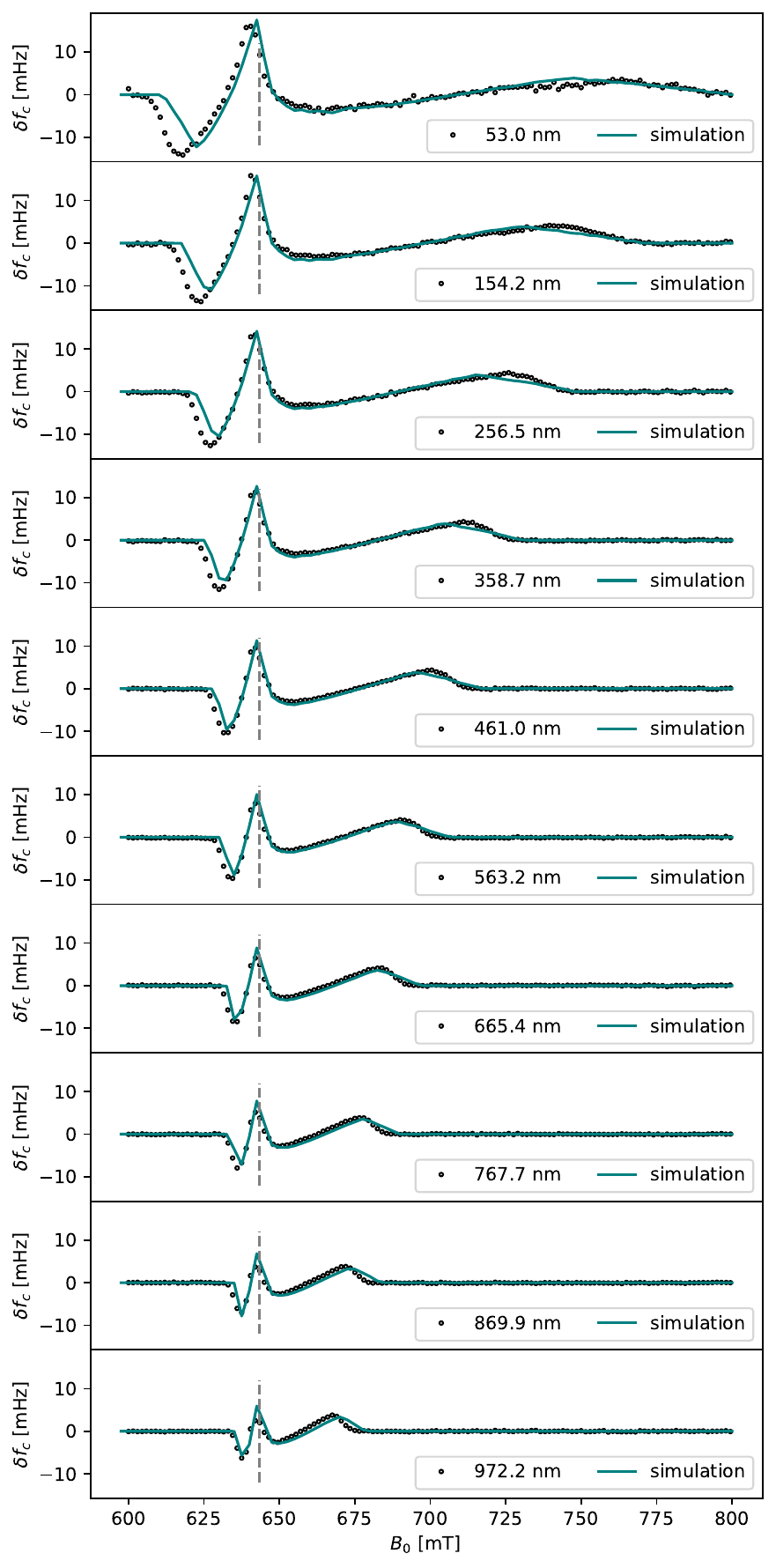}
\caption{Experiment data (circle) from a similar run as Ref.~\citenum{Moore2009dec} with experimental tip-sample separation indicated in the legend of each plot. The simulation for the \SI{200}{\nano\meter} uniform layer of tempamine-doped perdeuterated polystyrene used parameters in Table \ref{table:moore-exp-parameters}.  The gray dashed line is the external field at the Larmor frequency expected in the absence of the tip field.
} 
    \label{fig:Moore-fit}
\end{figure*}

In the experiment, the MW frequency was fixed at \SI{18.1}{\giga\hertz}, and the spectrum of the signal versus the external field was collected at different tip-sample separations. 
Unlike the fitting method in Ref.~\citenum{Moore2009dec}, we used a fixed temperature of \SI{11.0}{\kelvin} to simulate the spectrum using \texttt{mrfmsim.experiment.CermitTD} experiment class.
The experimental data (black dots) and the simulated data (teal lines) at various tip-sample separations are shown in Fig.~\ref{fig:Moore-fit}. 
The gray dashed line is the external field at which the magnet's Larmor frequency is the applied MW frequency.  
The simulated data is in good agreement with the experimental data, correctly predicting the two negative and two positive peaks. 
We would like to note that the fitting is adjusted with a \SI{-2.5}{\milli\tesla} $B_0$ shift for all spectra. 
The shift was adjusted to match the peak position at the Larmor frequency. 
This shift, we believe, compensates for a small error in tip magnet calibration, external field calibration, or small extra magnetic fields arising from magnetic parts in the apparatus.
The lineshape can be understood using Eq.~\ref{eq:moore-final}.

\subsubsection{Signal lineshape}
\begin{figure}[t]
    \centering
    \includegraphics[width=\columnwidth]{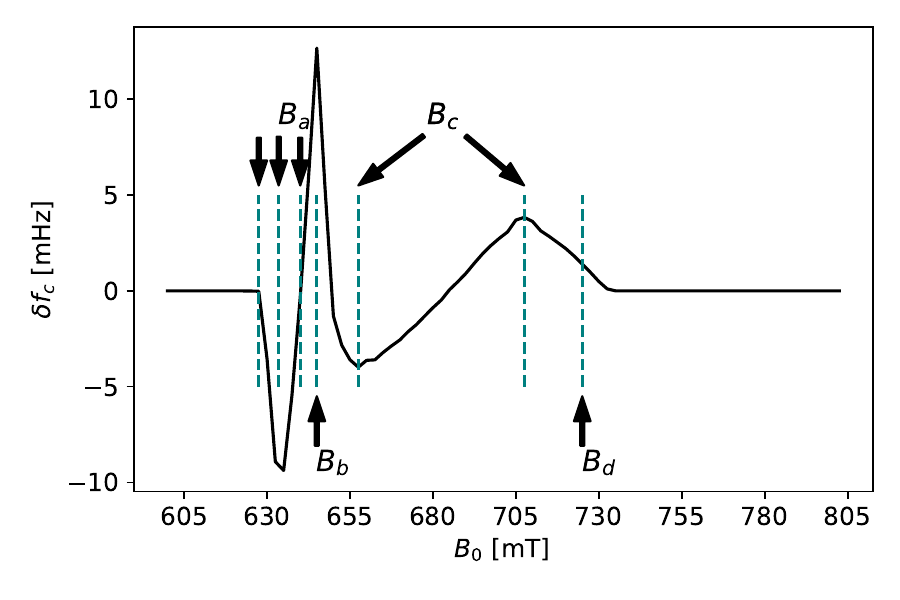}
\caption{Simulated signal of experiment in Ref.~\citenum{Moore2009dec} with a tip-sample separation of \SI{358.7}{\nano\meter}. The dashed vertical teal lines at $\mathrm{B}_a$ (\SI{627.5}{\milli\tesla}, \SI{633.5}{\milli\tesla}, and \SI{640}{\milli\tesla}), $\mathrm{B}_b$ (\SI{645}{\milli\tesla}), $\mathrm{B}_c$ (\SI{657.5}{\milli\tesla} and \SI{707.5}{\milli\tesla}), and $\mathrm{B}_d$ (\SI{725}{\milli\tesla}) indicates where the lineshapes are analyzed visually.}
    \label{fig:moore-lineshape}
\end{figure}

Let us analyze the simulated signal at a magnet tip-sample separation of \SI{358.7}{\nano\meter} in detail. 
Similarly to Sec.~\ref{subsubsec:Joni-lineshape}, the sensitive slice can be calculated using Eq.~\ref{eq:dMz_avg}, and the signal distribution can be determined by Eq.~\ref{eq:moore-final}.
We concentrated study on the signal at the specific fields $B\st{a}$, $B\st{b}$, $B\st{c}$, and $B\st{d}$ indicated in Fig.~\ref{fig:moore-lineshape}. 
The $B\st{a}$ features are at \SI{627.5}{\milli\tesla}, \SI{633.5}{\milli\tesla}, and \SI{640}{\milli\tesla}. 
The $B\st{b}$ field (\SI{645}{\milli\tesla}) is at the external field that produces the Larmor frequency that matches the applied MW frequency. 
The $B\st{c}$ fields (\SI{657.5}{\milli\tesla} and \SI{707.5}{\milli\tesla}) represent the signal's second negative and positive peaks. 
The $B\st{d}$ field (\SI{725}{\milli\tesla}) is close to the high-field end of the signal.

\begin{figure*}[!htbp]
    \centering
\includegraphics[width=0.8\textwidth]{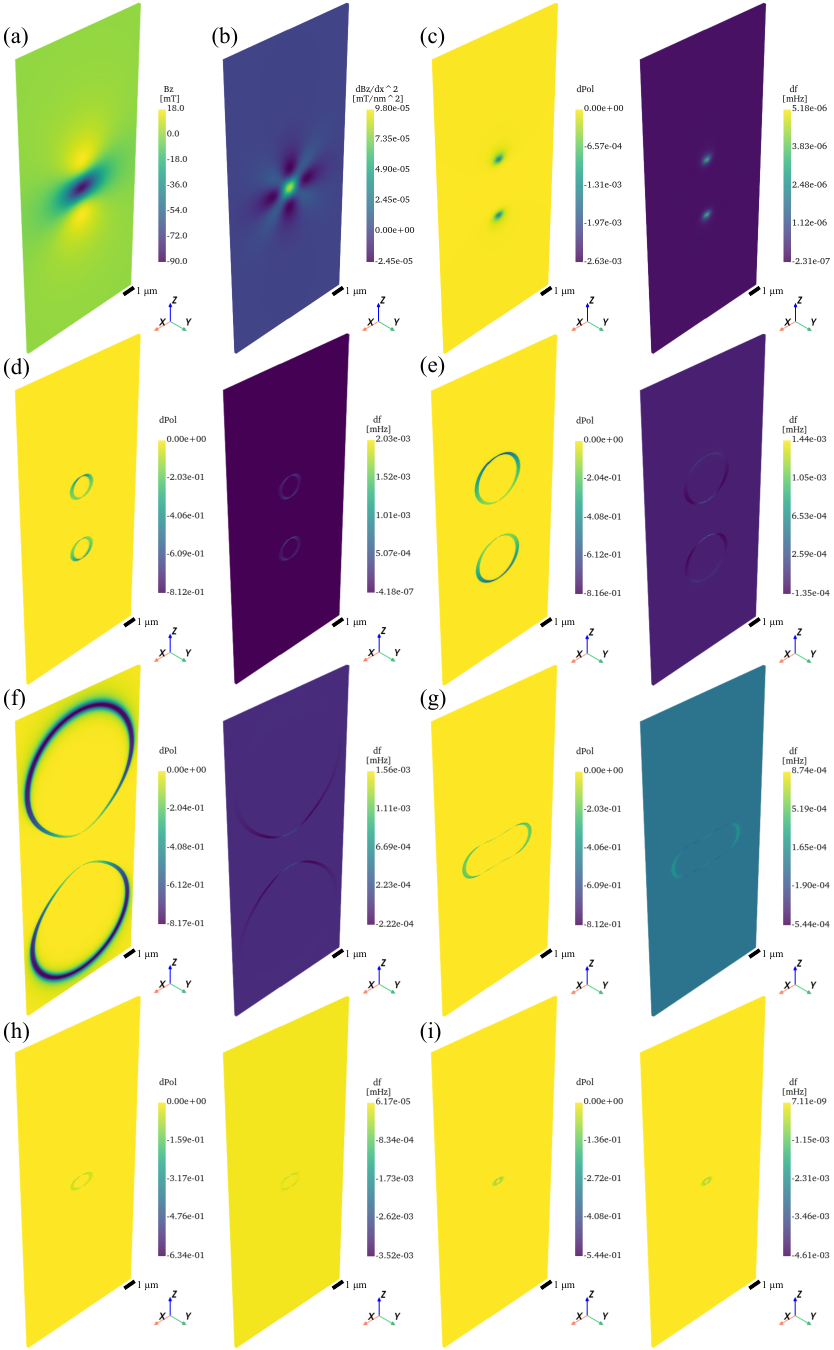}
\caption{Experiment lineshape render for Ref.~\citenum{Moore2009dec} at a \SI{358.7}{\nano\meter} tip-sample separation. The sample grid size is \SI{10}{\micro\meter} $\times$ \SI{0.2}{\micro\meter} $\times$ \SI{22}{\micro\meter}, with a \SI{20}{\nano\meter} step size. (a) Tip field $B_\mathrm{tip}$ overlayed with the sample film. (b) Tip field gradient $\partial^2 B_z^\mathrm{tip}/{\partial x}^2$ overlayed with the sample film. The sensitive slices and the signal distribution at the seven external fields, given in Fig.~\ref{fig:moore-lineshape} are shown in (c-i). Volumes are rendered using the \textit{PyVista} package.\citep{Sullivan2019may}} 
    \label{fig:moore-lineshape-render}
\end{figure*}

In Fig.~\ref{fig:moore-lineshape-render}, we render the tip magnetic field $B_\mathrm{tip}$, the field gradient $\partial^2 B_z^\mathrm{tip}/{\partial x^2}$, the sensitive slice, and the signal distribution at the selected features. 
Discrete colors are used for display purposes to achieve better contrast between positive and negative values. 
A spherical magnet in a uniform external field yields two positive lobes of tip field $B_\mathrm{tip}$, at the north and south poles of the magnet, and a torus-shaped negative lobe at the equator (Fig.~\ref{fig:moore-lineshape-render}(a)). 
Because the sample surface is aligned with the external field, some spins experience a positive tip field while others experience a negative tip field, with the negative field at the center.
The broad distribution of tip fields accounts for the large signal width.
Spins can experience a positive or negative $\partial^2 B_z^\mathrm{tip}/{\partial x^2}$, Fig.~\ref{fig:moore-lineshape-render}(b).
This observation accounts for the presence of a positive signal, negative signal, and points of zero signal where there is net cancellation of $\partial^2 B_z^\mathrm{tip}/{\partial x^2}$ over the sensitive slice. 

To explain the lineshape, we can start at the $B\st{b}$ field.
At $B\st{b}$, the spins that experience zero tip field are within the sensitive slice (Fig.~\ref{fig:moore-lineshape-render}(f)). 
With the external field lower than $B\st{b}$, at $B\st{a}$, the positive tip field lobes contribute to the sensitive condition until the slice no longer intersects with the sample (Fig.~\ref{fig:moore-lineshape-render}(c-e)). 
With the external field higher than $B\st{b}$, at $B\st{c}$ and $B\st{d}$, the negative torus contributes to the sensitive slice (Fig.~\ref{fig:moore-lineshape-render}(g-i)).
The signal sign depends on the sign and magnitude of the ${\partial^2 B_\mathrm{tip}^z / \partial x^2}$ distribution. 

The onset and offset of the signal are due to the tip field interacting with the sample, and therefore, they are dependent on the tip-sample separation. 
We can expect that increasing the tip-sample separation moves $B\st{a}$, $B\st{c}$, and $B\st{d}$ closer to the $B\st{b}$ field, while $B\st{b}$ remains the same. 
The $B\st{c}$ and $B\st{d}$ signals experience larger shifts due to the sharper gradient of the torus, which is more tip-sample separation dependent. 
The model explains the observation that, with the increasing tip-sample separation, the features in $B\st{a}$, $B\st{c}$, and $B\st{d}$ move closer to the $B\st{b}$ peak, while the $B\st{c}$ and $B\st{d}$ features shift faster with the change of tip-sample separation.
At lower tip-sample separation, the simulation underestimates the $B\st{a}$ negative peak, which suggests that spins nearest the tip experience a slightly higher tip field than the spherical-tip approximation predicts. 
The difference suggests the magnet surface is not smooth, which creates an irregularity in the local magnetic field at low tip-sample separation.
This observation is in agreement with scanning electron microscope images in Ref.~\citenum{Moore2009dec} that reveal a bumpy magnet surface at high magnification.

The simulation model used in Ref.~\citenum{Moore2009dec} assumed that spins in the resonance condition saturate as if radiation is applied in the steady state. 
The assumption was incorrect, as shown in Eq.~\ref{eq:Mz(tau0)}; a spin, in general, does not have enough time to saturate during the cantilever sweep. 
As a result, the Ref.~\citenum{Moore2009dec} simulation overestimated the signal at the $B\st{c}$ fields.
Furthermore, the simulation published in Fig.~7 of Ref.~\citenum{Moore2009dec} approximated the cantilever sweep by moving the cantilever in increments of the grid spacing. The resonance offset was chosen where saturation is greatest. This algorithm is much more sensitive to grid spacing than Eq.~\ref{eq:Mz(tau0)}, and results in discontinuities in the calculated saturation.

While the accidental compensation of signal error from an incorrect signal model and a too-coarse simulation grid fit the data at selected tip-sample separations, the simulation overall did not fit data at all tip-sample separations very well, and gave an erroneous estimate of $B_1$. 
We include a more detailed discussion on the fitting limitation of Ref.~\citenum{Moore2009dec} algorithm in Sec.~\ref{appendix:moore-discussion}.
With the Ref.~\citenum{Boucher2023sep} signal model, for the $B\st{c}$ features, the spins that contribute to the signal experience a higher field gradient, and as a result, they are saturated less in comparison with the features in $B\st{a}$ and $B\st{b}$. As a result, the simulation with the new algorithm fits the experimental line shape very well, shown in Fig.~\ref{fig:Moore-fit}, matching both the bulk features at $B\st{a}$ and $B\st{b}$ fields and local features at $B\st{c}$ and $B\st{d}$ fields.
 \section{Conclusions}

In this paper, we have presented \textit{mrfmsim}, a modular, extendable, readable, and well-tested package for simulating MRFM experiments. 
By building on \textit{mmodel}, the \textit{mrfmsim} package enables much faster development, facilitates collaboration, and produces reproducible simulations. 
The workflow is designed for a graduate research setting and creates a shallower learning curve for collaboration. 
The package's modular design encourages the community to add experimental models and features to \textit{mrfmsim}. 
Overall, \textit{mrfmsim} has shown a 20-fold speed up when compared to previous simulations, and has led to new spin-physics discoveries through comparison to experiment.\citep{Boucher2023sep} 
In this paper, we simulated, analyzed, and reproduced the data of the nuclear spin noise experiment in Ref.~\citenum{Longenecker2012nov} and the CERMIT force-gradient electron spin experiment in Ref.~\citenum{Moore2009dec}. The \textit{mrfmsim} package revealed an error in the simulations reported in Ref.~\citenum{Moore2009dec}, which we have corrected here. 
We used \textit{mrfmsim-plot} to visualize resonant spins and rationalize the observed lineshape in both experiments.
Simulating signals is a barrier to beginning MRFM experiments, and \textit{mrfmsim} significantly lowers this barrier.
We believe \textit{mrfmsim} can accelerate ongoing experiments by enabling a faster development cycle for experimental design and analysis.
Finally, \textit{mrfmsim} can serve as an example of how to design a simulation package for a research field undergoing continuous development.

The version 0.4.1 of the \textit{mrfmsim} package is open source and published to PyPI at \url{https://pypi.org/project/mrfmsim/0.4.1/}.
Detailed documentation for \textit{mrfmsim} is available at \url{marohn-group.github.io/mrfmsim-docs/}. The plugin modules are available at \url{https://github.com/Marohn-Group/mrfmsim-cli} (\texttt{mrfmsim-cli}), \url{https://github.com/Marohn-Group/mrfmsim-unit} (\texttt{mrfmsim-unit}), \url{https://github.com/Marohn-Group/mrfmsim-plot} (\texttt{mrfmsim-plot}), and \url{https://github.com/Marohn-Group/mrfmsim-yaml} (\texttt{mrfmsim-yaml}). The code used for the analysis and figure generation in this paper is available at \url{https://github.com/Marohn-Group/mrfmsim-examples}.
The projects are open-sourced under the 3-Clause BSD License.
 \section{Acknowledgments}

Research in this paper was supported by Cornell University, the Army Research Office under Award Number W911NF-17-1-0247, and the National Institute Of General Medical Sciences of the National Institutes of Health under Award Number R01GM143556.
 
\clearpage
\appendix
\renewcommand{\thetable}{A.\arabic{table}}
\setcounter{table}{0}

\renewcommand{\thefigure}{A.\arabic{figure}}
\setcounter{figure}{0}

\section{Development of steady-state algorithm for the CERMIT experiments}\label{appendix:moore-discussion}

In Sec.~\ref{subsection: mrfmsim-moore-tdo}, we discussed that the model Ref.~\citenum{Moore2009dec} used to approximate the spin saturation under the microwave radiation was incorrect due to the incorrect steady state assumption. We presented an updated simulation model that accounts for spin relaxation, showing good agreement with the experimental data. In this appendix, we discuss the grid-size dependence of the original algorithm used in Ref.~\citenum{Moore2009dec} and present an updated algorithm, initially presented in Ref.~\citenum{Isaac2018feb}, under the steady-state assumption.

As shown in Eq.~\ref{eq:moore-final}, given that the magnet tip size is large, the CERMIT signal resulting from the magnetization change is
\begin{equation}\label{eq:cermit-final}
    \delta f = -\
\frac{\sqrt{2}f_\mathrm{c}}{2\pi k_\mathrm{c}}\sum_j \Delta M_z(\bm{r}_j) \frac{\partial ^2 B_z ^\mathrm{tip} (\bm{r}_j)}{\partial x^2}
\end{equation}
where $\Delta M_z(\bm{r}_j)$ is the change in polarization for spin at position $\bm{r}_j$, $\partial B_z ^\mathrm{tip} (\bm{r}_j)/ \partial x^2$ is the second derivative of the magnet tip field in the $x$ direction, $f_\mathrm{c}$ is the cantilever frequency, $k_\mathrm{c}$ is the cantilever spring constant, and the $\sqrt{2}/\pi$ factor accounts for the root-mean-square lock-in output of the demodulated frequency change.\citep{Lee2012apr}
The relative change in magnetization from the steady-state solution of the Bloch equation for each spin is 

\begin{equation}\label{eq:ss}
    \Delta M_z(\bm{r}_j) = - \frac{\gamma_\mathrm{e}^2 B_1^2  T_1 T_2}{1 + \gamma_\mathrm{e}^2 \Delta B_0(\bm{r}_j) ^2T_2^2 + \gamma_\mathrm{e}^2 B_1^2 T_1 T_2}M_0
\end{equation}
where  $\Delta B_0(\bm{r}_j) = B\st{ext} + B_z\sr{tip}(\bm{r}_j) - 2\pi f\st{mw}/\gamma_\mathrm{e}$ is the resonance offset, $B_z\sr{tip}(\bm{r}_j)$ is the tip magnetic field at the position $\bm{r}_j$, $f\st{mw}$ is the applied MW frequency, and $\gamma_\mathrm{e}$ is the electron gyromagnetic ratio.\citep{Nguyen2018feb}

For large magnet tips, the cantilever motion is small compared to the tip size, shown in Table \ref{table:moore-exp-parameters}. 
Therefore, we assume that the $\partial B_z ^\mathrm{tip} (\bm{r}_j)/ \partial x^2$ term in Eq.~\ref{eq:cermit-final} stays constant throughout the cantilever motion. As a result, the difference in the frequency signal, $ \delta f$, originated from the difference in the change in magnetization, $\Delta M_z$.  The $ \delta f$ and subsequently $\Delta M_z$ reach the maximum when the resonance offset $\Delta B_0$ reaches zero. To approximate the resonance offset, additional grid points --- discretized by the grid step size --- were used to represent the change in resonance offset during the cantilever motion. A final value of resonance offset was chosen based on algorithms. For simplicity, we show an example in Table \ref{table:algo-example} where we approximate the motion using a 5-point array, $P_0$ - $P_4$, and we populate the array at each spin location, $x_0$ - $x_4$, with linear resonance offset values. In the example, the spin experiences the difference in resonance offset as if the magnetic tip field is moving to the right from $P_0$ to $P_4$.

\begin{figure}[t]\label{fig:algo-grid-compare}
    \centering
\includegraphics[width=\columnwidth]{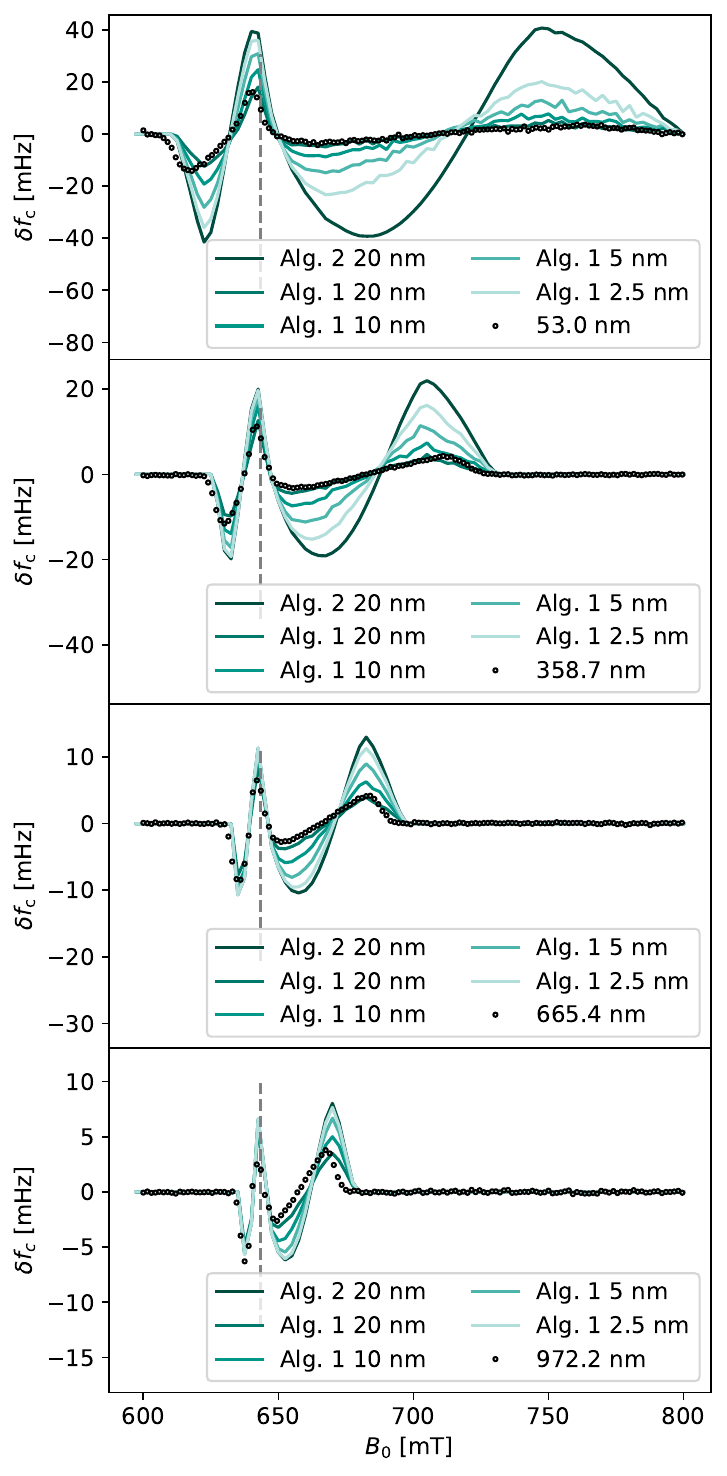}
\caption{Simulation for the Ref.~\citenum{Moore2009dec} experiment with different grid sizes at different tip-sample separations for Algorithm 1 (Alg.~1) and Algorithm 2 (Alg.~2). We select four tip-sample separation signals from Fig.~\ref{fig:Moore-fit}. The grid size for the simulation is \SI{10}{\micro\meter} $\times$ \SI{0.2}{\micro\meter} $\times$ \SI{22}{\micro\meter}, and the step size in the $y$ and $z$ directions are \SI{20}{\nano\meter}. At each tip-sample separation, five simulations are performed: Alg.~1 at \SI{20}{\nano\meter}, \SI{10}{\nano\meter}, \SI{5}{\nano\meter}, and \SI{2.5}{\nano\meter} $x$ step sizes, and Alg.~2 at \SI{20}{\nano\meter} $x$ step size. } 
    
\end{figure}

The algorithm used in Ref.~\citenum{Moore2009dec} (Algorithm 1) approximates the resonance offset for each spin by its minimum absolute value throughout the cantilever motion. 
When the resonance offset value crosses zero during the cantilever motion (the value changes signs), the spins are assumed to be fully saturated. However, the saturation is not reflected in the final approximated resonance offset values, resulting in an underestimation. The underestimation scales with the grid step size --- the larger the step size, the more the underestimation. The effect can be seen in the Alg.~1 plots of the Fig.~\ref{fig:algo-grid-compare}. The grid step size used in  Ref.~\citenum{Moore2009dec} was \SI{20}{\nano\meter}, which fits the experimental data well due to the underestimation. However, when the grid size is reduced, the simulation no longer agrees with the experimental data. 

 An improved approximation discussed in Ref.~\citenum{Isaac2018feb} is shown as Algorithm 2 in Table \ref{table:algo-example}. In this algorithm, we assume that if the resonance offset changes signs during the cantilever motion, the spin is fully saturated with the final resonance offset value of zero (location $x_0$ - $x_3$ in Table ~\ref{table:algo-example}). In comparison, Algorithms 1 and 2 have the same result when the resonance offset does not change signs ($x_4$ in Table \ref{table:algo-example}). However, if the resonance offset crosses zero during the cantilever sweep ($x_0$ - $x_3$ in Table ~\ref{table:algo-example}), Algorithm 2 is able to simulate fully saturated spins with resonance offset values of 0. 
In Fig.~\ref{fig:algo-grid-compare}, we show the comparison of Algorithms 1 and 2 using parameters from the Table.~\ref{table:moore-exp-parameters} at different grid step sizes. The $B_1$ value used is \SI{0.29}{\micro\tesla}, same as Ref.~\citenum{Moore2009dec}.
 We observe that at a \SI{20}{\nano\meter} step size, Algorithm 1 fits the experimental data well, as suggested in Ref.~\citenum{Moore2009dec}. However, as the step size decreases, Algorithm 1 deviates from the experimental data. In comparison, Algorithm 1, with a \SI{2.5}{\nano\meter} step size, still underestimates the signal relative to Algorithm 2 at a \SI{20}{\nano\meter} step size. Simulated signals with step sizes of \SI{10}{\nano\meter}, \SI{5}{\nano\meter}, and \SI{2.5}{\nano\meter} are nearly identical to those with a \SI{20}{\nano\meter} step size using Algorithm 2.

In summary, under the assumption that the spin reaches the steady state saturation condition when the radiation is applied, Algorithm 1 underestimates the true signal and leads to the incorrect conclusion in Ref.~\citenum{Moore2009dec}. Here, we showed an improved Algorithm 2 that correctly approximates the signal under the assumption.
The Algorithm 2 is included in the \textit{mrfmsim} package as \texttt{mrfmsim.experiments.CermitESR} experiment class.

\begin{table}[htbp]\label{table:algo-example}
\centering
\caption{Algorithm comparison of the change in resonance offset, $\Delta B_0$, calculated from two different saturation algorithms. The location $x_{0, 1, 2, 3, 4}$ represents the locations of the spins on the left side of the grid. The cantilever moves from the left ($P_0$) to the right ($P_4$) in the $x$ direction. For a grid step size of \SI{20}{\nano\meter}, the cantilever peak-to-peak value is \SI{80}{\nano\meter} for this example. The $\Delta B_0$ values are represented with arbitrary values.  In Algorithm 1, the minimum absolute value during the full sweep is used for the final value. In Algorithm 2, if the resonance offset crosses zero, representing a full saturation, the final $\Delta B_0$ is 0. Otherwise, the minimum absolute value during the full sweep is used for the final value.}
\begin{tabular}{l r r r r r r}
\hline
Cantilever Position  & $x_0$ & $x_1$ & $x_2$ & $x_3$ & $x_4$ & ... \\
\hline
$P_0$   & 1 & 3 & 5 & 7& 9 & ... \\
$P_1$    & -1 & 1 & 3 & 5 & 7 & ...\\
$P_2$   & -3 & -1 & 1 & 3 & 5 & ...  \\
$P_3$     & -5 & -3 & -1 & 1 & 3 & ... \\
$P_4$   & -7  & -5 & -3 & -1 & 1 & ...\\
\hline
Algorithm 1 (Ref.~\citenum{Moore2009dec})  & 1 & 1 & 1 & 1 & 1 & ...  \\
Algorithm 2 (Ref.~\citenum{Isaac2018feb})  & 0 & 0 & 0 & 0 & 1 & ...  \\
\hline
\end{tabular}
\end{table}

\clearpage
\bibliography{bib/reference_static.bib}

\end{document}